# Highly linear polarized emission at telecom bands in InAs/InP quantum dot-nanowires by geometry tailoring

*Ali Jaffal[1,2], Philippe Regreny[1], Gilles Patriarche[3], Michel Gendry[1] and Nicolas Chauvin*[2]*

[1] Univ Lyon, CNRS, Ecole Centrale de Lyon, INSA Lyon, Université Claude Bernard Lyon 1, CPE Lyon, CNRS, INL, UMR5270, 69130 Ecully, France

[2] Univ Lyon, CNRS, INSA Lyon, Ecole Centrale de Lyon, Université Claude Bernard Lyon 1, CPE Lyon, INL, UMR5270, 69621 Villeurbanne, France

[3] Université Paris-Saclay, CNRS, Centre de Nanosciences et de Nanotechnologies - C2N, 91120, Palaiseau, France

Email: nicolas.chauvin@insa-lyon.fr

ABSTRACT

Nanowire (NW)-based opto-electronic devices require certain engineering in the NW geometry to realize polarized-dependent light sources and photodetectors. We present a growth procedure to produce InAs/InP quantum dot-nanowires (QD-NWs) with an elongated top-view cross-section relying on the vapor-liquid-solid method using molecular beam epitaxy. By interrupting the rotation of the sample during the radial growth sequence of the InP shell, hexagonal asymmetric (HA) NWs with long/short cross-section axes were obtained instead of the usual symmetrical shape. Polarization-resolved photoluminescence measurements have revealed a significant influence of the asymmetric shaped NWs on the InAs QD emission polarization with the photons being mainly polarized parallel to the NW long cross-section axis. A degree of linear polarization (DLP) up to 91% is obtained, being at the state of the art for the reported DLP values from QD-NWs. More importantly, the growth protocol herein is fully compatible with the current applications of HA NWs covering a wide range of devices such as polarized light emitting diodes and photodetectors.

INTRODUCTION

Integrating III-V semiconductor nanowires (NWs) on silicon substrates has been widely studied to achieve telecom band NW lasers[1] and single photon sources[2], light emitting diodes[3], photovoltaics cells[4] or photodetectors.[5] Tailoring the NW shape is a key issue to optimize the device efficiency. For instance, introducing a taper along the NW length reduces the beam divergence of single photon sources[6], and improves the broadband, angle insensitive and absorption efficiency in arrays of NWs for solar cells or photodetectors.[7]

However, in the case of bottom-up NWs, the in-plane shape is usually a hexagonal one as a consequence of the growth of NWs with a wurtzite (WZ) crystallographic phase along the [0001] axis or a zinc blende (ZB) phase along the [111] axis. Tailoring the NW shape to get in-plane asymmetric NWs (also called elliptic NWs) would pave the way to linearly polarized sensitive devices. For example, arrays of asymmetric NWs could be used for polarization-resolved imaging[8] or polarized light emitting diodes.[9] At the single NW level, an asymmetric NW can be used to get a single photon source with a deterministic polarized state.[10] In polarization-resolved NW devices, no polarization filters are required. Therefore, the unwanted polarized light is not discarded by the filter, which increases the theoretical efficiency of such NW devices.

In the field of linearly polarized light sources with quantum dot-nanowires (QD-NWs), two methods of fabrication are reported in literature: The first one is the top-down fabrication of elliptic NWs using etching.[11] However, the number and position of the QDs per NW are not controlled with this procedure. Another strategy is the bottom-up growth of hexagonal-asymmetric (HA) NWs using of Metal Organic Vapour Phase Epitaxy (MOVPE) with the assistance of patterned elongated nanoscale openings within a silicon dioxide mask on a GaAs substrate.[10] This method presents a strong limitation because of the unavoidable growth of large QDs inside the NWs due to the elongated NW cross-section. This large QD, having the same diameter as the NW one along the long cross-section axis, results in a weak carrier confinement inside the QD with a small energy separation between the confined states[12] and a broad emission linewidth.[13] Thus, realizing HA NWs using a bottom-up approach without any substrate patterning and with a small QD located on the growth axis of the NWs is still a remaining challenge.

In this work, we experimentally demonstrate a growth method to realize HA InAs/InP QD-NWs grown on Si(111) substrates using the vapor-liquid-solid (VLS) method by solid-source molecular beam epitaxy (ss-MBE). By using two growth steps consisting of rotating the sample during the axial growth of the InAs/InP QD-NWs while rotation interruption is implemented during the radial growth of the InP shell, we enable the transformation of the hexagonal-symmetric (HS) NWs into HA NWs. Polarization resolved photoluminescence (PL) together with Finite Difference Time Domain (FDTD) simulations fully confirm the predicted performances with the QD emission being polarized parallel to the long NW cross-section axis and a degree of linear polarization (DLP) value as high as 91%.

DESIGN OF THE HEXAGONAL ASYMMETRIC QD-NWs

In a standard symmetric QD-NW, the NW diameter ($D_{NW}$) has to be tuned to efficiently couple the QD emission to the fundamental $HE_{11}$ NW guided mode. In high refractive index NWs such as GaAs and InP NWs the optimized diameter is in the order of $0.25 \times \lambda$, where $\lambda$ is the QD emission wavelength.[2] This type of symmetric NW, operating in the single mode regime, supports actually two weakly nondegenerate fundamental optical modes ($HE_{11x}$ and $HE_{11y}$) being strongly confined and orthogonally polarized. This means that, if a single QD is embedded along the NW growth axis, the photons emitted by the two QD orthogonally exciton polarized bright states will funnel equally into their respective optical mode and therefore the QD emission will be unpolarized. To transform this NW into a polarized light source, the diameter has to be reduced along one axis (the short axis) to expel the concerned mode in the air cladding and to inhibit the radiative lifetime along this axis. In a cylindrical NW it is known that this condition is reached when $D_{NW} < 0.15 \times \lambda$.[14] Therefore, the short axis has to be smaller than 200 nm for a QD emitting in the telecom band. In Figure 1a and b, we show the optical mode profiles $M_x$ and $M_y$ in the directions parallel and perpendicular to the NW long cross-section axis of a HA NW with 420/170 nm long/short axes, respectively. As the modal effective refractive index, $n_{eff}$, depends on the NW diameter and shape[15], the $n_{eff}$ for $M_x$ is higher than that of $M_y$ with values of 1.7 and 1, respectively. As a result, the optical mode $M_x$ is mainly confined inside the NW whereas the $M_y$ mode is delocalized in the air cladding. With this geometry, the PL emission from the QD will be efficiently funneled into the $M_x$ mode and a linearly polarized PL emission along the x-axis should be detected.

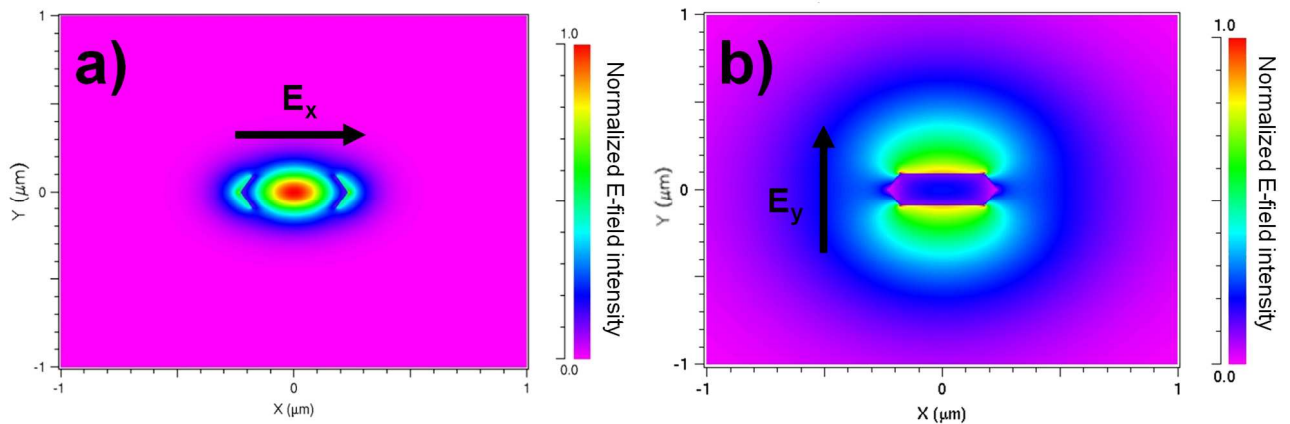

**Figure 1 -** FDTD simulations showing the normalized electric field profiles of the two optical modes $M_x$ (a) and $M_y$ (b) at a 1.4 µm wavelength.

GROWTH OF THE HA InAs/InP QD-NWs

The growth of InAs/InP QD-NWs was initiated with the in-situ formation of In-Au droplets by the In and Au co-evaporation during 60 sec at 500°C as a catalyst for the VLS growth of the NWs.[16] We have chosen an In/Au beam equivalent pressure (BEP) ratio for the In-Au droplet formation equal to 58 with the In BEP = $2.5 \times 10^{-7}$ torr corresponding to a growth rate of 0.2 ML/sec for an InP two-dimensional (2D) layer and the Au BEP = $4.2 \times 10^{-9}$ torr. This is followed by growing pure WZ InAs/InP QD-NWs at 420°C/380°C with a V/III BEP ratio = 20 for the InAs QD and InP NW under continuous rotation. With these growth conditions, ultra-low NW density < 0.1 µm$^{-2}$ was achieved (Figure 2a). More details on this growth step can be found in Ref 17. The InP NWs after this growth step have a diameter in the 50-60 nm range and 1.1-1.2 µm length (Figure 2b) and the InAs QD position is estimated to be 300 nm below the catalyst droplet at the upper part of the NWs. The QD, in Figure 2c, has a diameter D and a height H in the range of 23-28 nm and 3-5 nm, respectively, thus with H/D ≈ 0.1, and with a hexagonal symmetry.[2,17] This procedure ensures the on-axis positioning of the QDs inside the NWs. Moreover, the H/D ≈ 0.1 ratio guaranties a strong heavy-hole ground state in the valence band.[18,19,20] This means that the exciton dipole is oriented perpendicularly to the NW growth axis and that the exciton transition can be described by two orthogonally-polarized bright states which are split by the fine-structure splitting. This splitting is expected to be very small, in the µeV order, for InAs/InP QD-NWs grown along the [0001] or [111] axis.[21,22]

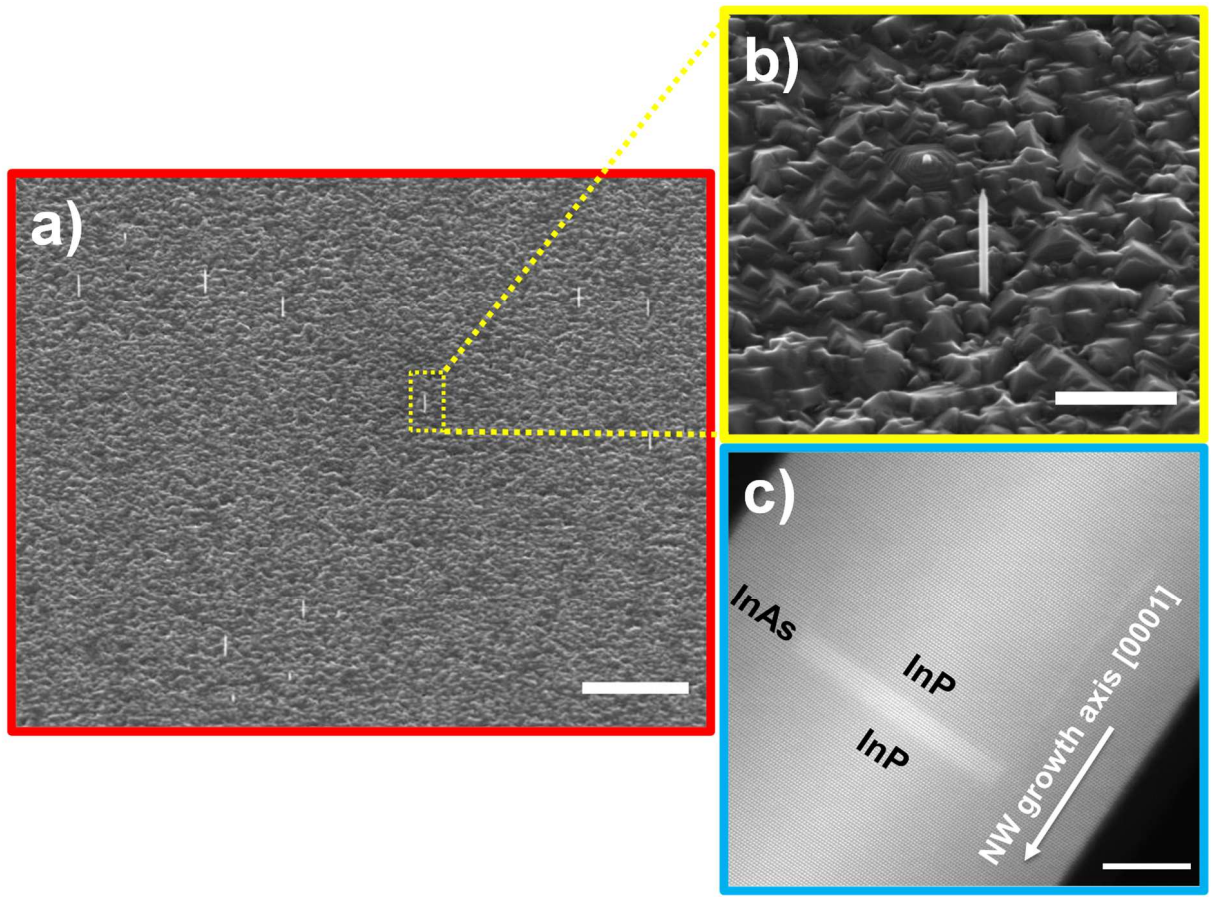

**Figure 2** - (a,b) Tilted view (45°) SEM images of the InAs/InP QD-NWs showing the low NW density (< 0.1 µm$^{-2}$). c) High-angle annular dark-field scanning transmission electron microscopy image of an InAs QD inside an InP NW showing the on-axis positioning of the QD in the NW with a QD aspect ratio H/D ≈ 0.1. Scale bars in (a), (b) and (c) are 5 µm, 1 µm and 10 nm, respectively.

To transform the HS NWs to HA NWs, a second growth step was added by growing a radial InP shell using specific growth conditions.[23] After the axial growth step, the growth was stopped by closing the In shutter while keeping the sample under P$_2$ flux and the sample temperature was decreased to 340°C during 3 min to then favor the radial growth over the axial growth.[24] Then, the sample rotation was stopped and the hexagonal NWs were aligned with the In cell along the [11-20] direction (equivalent to the cubic direction [1-10]) as illustrated in Figure 3a, with the use of the reflection high energy electron diffraction (RHEED) pattern (Figure 3b). To grow the InP shell on the chosen direction, we have rotated the sample by an angle of 130° which is the angle between the positions of the RHEED screen and the In cell (Figure 3c). As a result, a [11-20] direction and thus a corner of the hexagonal shape of the NWs is therefore perfectly aligned with the In cell flux and the first growth stage of the InP shell can be initiated by opening the In shutter. During the radial growth (see the schematic illustration in Figure 3d), we have used a high P$_2$ BEP equal to 2x10$^{-5}$ torr so that we ensure phosphorus-rich growth conditions to obtain defect-free WZ NWs and to limit the diffusion of the In adatoms towards the Au catalyst thus favoring radial growth over the axial growth. The InP shell was grown using an In BEP equal to 5x10$^{-7}$ torr and 10$^{-6}$ torr corresponding to an In flux equal to 0.5 ML/sec and 1 ML/sec, respectively. After 14 – 20 min of InP shell growth time, the growth was stopped during 1 min by closing the In shutter and the sample was rotated by 180° to grow the second stage of the InP shell. By re-opening the In shutter, the second InP shell growth stage was started on the opposite side of the NWs with the exact growth conditions used for the first InP shell growth stage. After this radial growth step, the sample rotation was restarted immediately with the opening of the Ga shutter during 90 sec at 340°C, to passivate the NWs by a Ga$_{0.15}$In$_{0.85}$P outer shell.

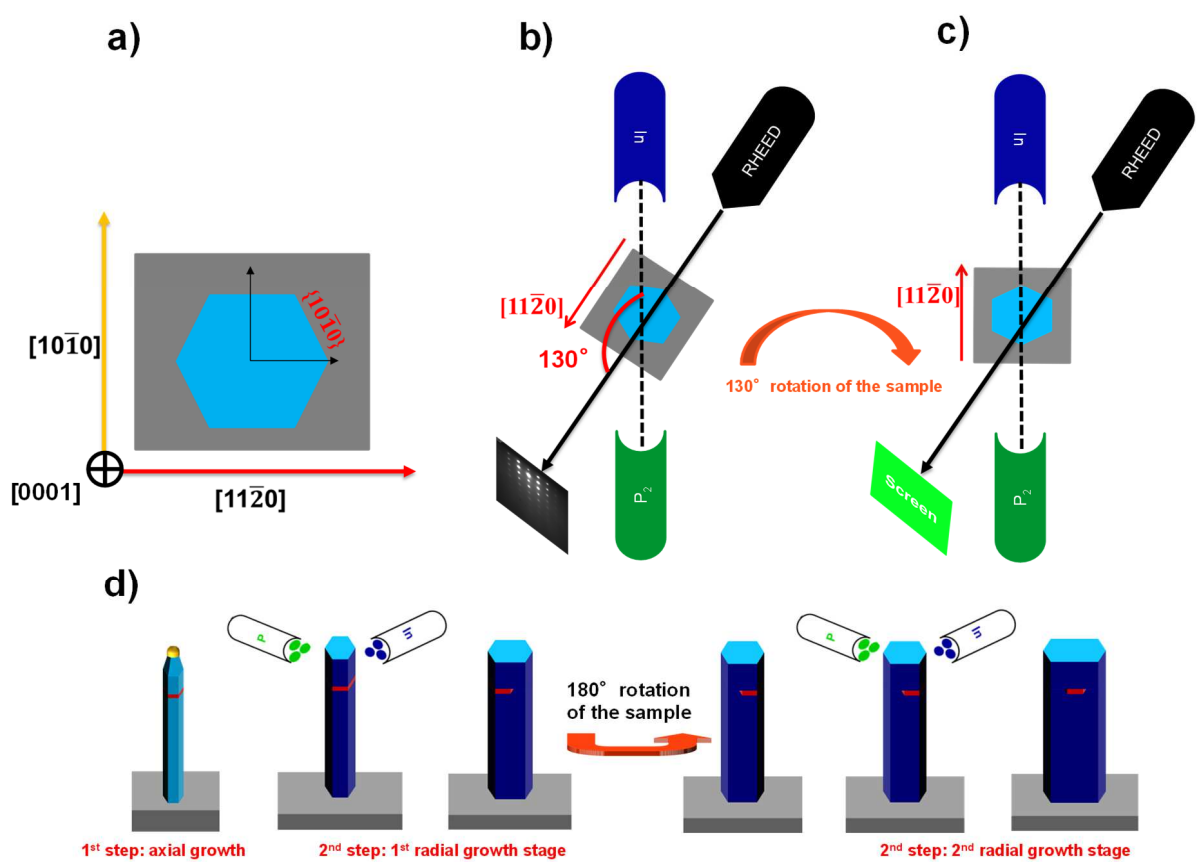

**Figure 3** - a) Schematic representation of a hexagonal symmetric WZ NW in top-view showing the crystallographic directions and {10-10} facets of the NW. (b,c) Schematic representation of the procedure to align a [11-20] direction of the NWs with the In cell flux using the RHEED pattern. d) Schematic representation of the growth steps from left to right that were performed to reach HA NWs. As an example, the In cell is aligned with the [11-20] direction of the NWs. To better explain the shape (in light blue) evolution of the NW cross-section, the catalyst droplet and the NW top have been suppressed in the 2$^{nd}$ step of the growth.

The top-view SEM images in Figure 4 show the different HS (a) and HA (b-d) InAs/InP QD-NWs we have obtained as a function of the procedure used for the growth of the radial InP shell (the growth conditions are summarized in Table 1). The NWs with HS cross-section in Figure 4a were obtained without introducing any rotation interruption during the radial growth of the InP shell. On the contrary, under different growth conditions in Figure 3(b-d), we can observe NWs with an elongated cross-section: after the axial growth, the obtained NWs with 50-60 nm in diameter have grown in an asymmetric manner during the growth of the InP shell. The growth in the chosen [11-20] direction has resulted in an elongation in this direction from 50-60 nm to 265-480 nm (long cross-section axis). On the other hand, the NW shape evolution in the direction perpendicular to the [11-20] direction is limited with a slower radial growth rate leading to a smaller increase from 50-60 nm to 155-170 nm (short cross-section axis). This difference in the radial growth rate is a result of the implemented growth procedure i.e. without rotating the sample during the radial growth, and thus asymmetric NWs are successfully obtained with the radial growth taking place mainly along the direction parallel to the [11-20] direction aligned with the In cell flux. The slight increase in the NW size in the direction perpendicular to the [11-20] direction is partly explained by the "radial" diffusion of the In adatoms on the NW sidewalls which is not completely suppressed at 340°C. By controlling either the growth time (14 min or 20 min) or the In flux (0.5 ML/sec or 1 ML/sec), we have successfully modified the InP NW long/short cross-section axes with values ranging from 155/265 nm (Figure 4b) to 170/480 nm (Figure 4d).

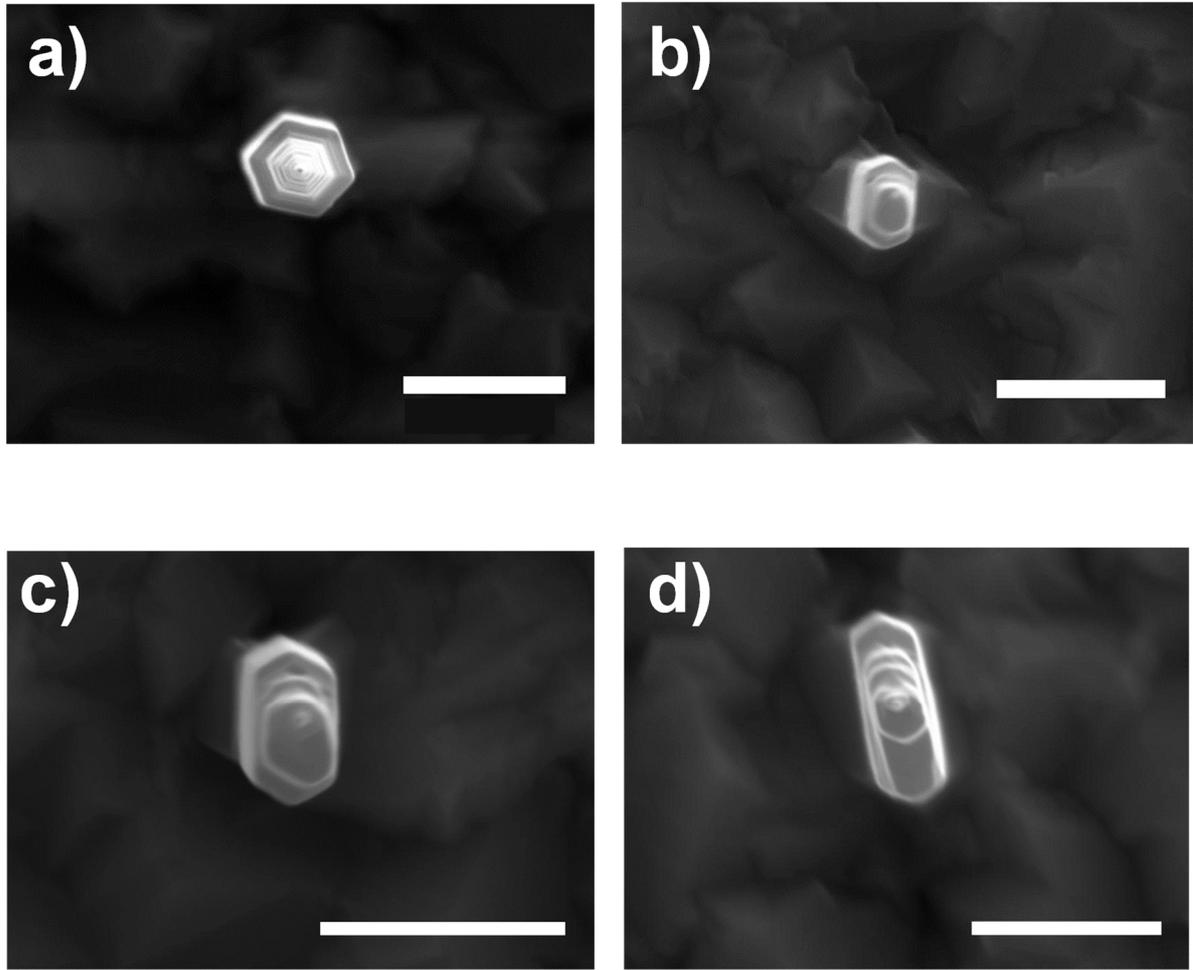

**Figure 4** – Top view SEM images of InAs/InP QD-NWs with different cross-section geometries. (a) hexagonal-symmetric cross-section obtained by the continuous rotation of the sample during the radial growth of the InP shell. (b-d) Hexagonal-asymmetric cross-section with different long/short cross-section axis dimensions obtained by interrupting the rotation of the sample during the radial growth and tailored by the InP shell growth time-In flux: b) 14 min – 0.5 ML/sec, c) 20 min – 0.5 ML/sec, d) 14 min - 1 ML/sec, on each side of the NWs parallel to the [11-20] direction. Scale bars in the inset and (a-d) are 50 and 500 nm, respectively.

| Figure 4 | Radial growth temperature (°C) | Growth time (min) | In flux (ML/sec) | Long/Short axes (nm) |
|---|---|---|---|---|
| (b) sample A | 340 | 14 | 0.5 | 265/155 |
| (c) sample B | 340 | 20 | 0.5 | 320/165 |
| (d) sample C | 340 | 14 | 1 | 480/170 |

**Table 1 –** Summary of the different growth conditions used to grow the InP shell for the HA InAs/InP QD-NWs shown in Figure 4.

In Figure 5, we show a tilted view SEM image of HS (a) and HA (b) InAs/InP QD-NWs around 4 µm in length. Contrary to the NW in Figure 5a which possesses a uniform tapering morphology, the upper part of the NW in Figure 5b exhibits a non-uniform and broad tapering section. It is well known that the NW conical tapering section is formed by the axial growth taking place simultaneously with the radial growth of the InP shell.[24] In our previous work[2], we have managed the growth to obtain a full control over the NW taper angle by adapting the InP shell growth temperature with the upper part of the NWs showing a uniform and smooth tapering angle (Figure 5a). Such a tapering mechanism was realized thanks to the balanced axial/radial growth rates occurring while the sample is rotating. On the contrary, when the sample rotation was stopped to grow the InP radial shell, an inhomogeneous NW taper has appeared. This is, most probably, due to the difference in the axial growth rate arising from the modification in the $P_2$ supply received by the NW side facets. First, we have to remember that the In and $P_2$ cells are 180° away from each other and therefore there is a difference in the $P_2$ flux received by the NW side facets facing the In and $P_2$ cells, In and $P_2$ BEP being equal to $1.45 \times 10^{-6}$ torr and $2 \times 10^{-5}$ torr, respectively. During the InP radial growth and at the NW side facets receiving a low $P_2$ supply (facing the In cell), the In adatom diffusion towards the catalyst droplet is higher promoting faster axial growth with In adatoms being the main source for the axial growth of Au-catalyzed In-based NWs.[25] Keeping in mind that the In adatom radial diffusion on the NW side walls towards the NW side facets receiving high $P_2$ supply (facing the $P_2$ cell) is not completely suppressed. Therefore, InP axial/radial growths could take place where the radial growth is more favored than the axial growth due to the elevated $P_2$ BEP. This would result in a difference in the axial growth rates at the opposite NW side facets and therefore would induce inhomogeneous NW taper.

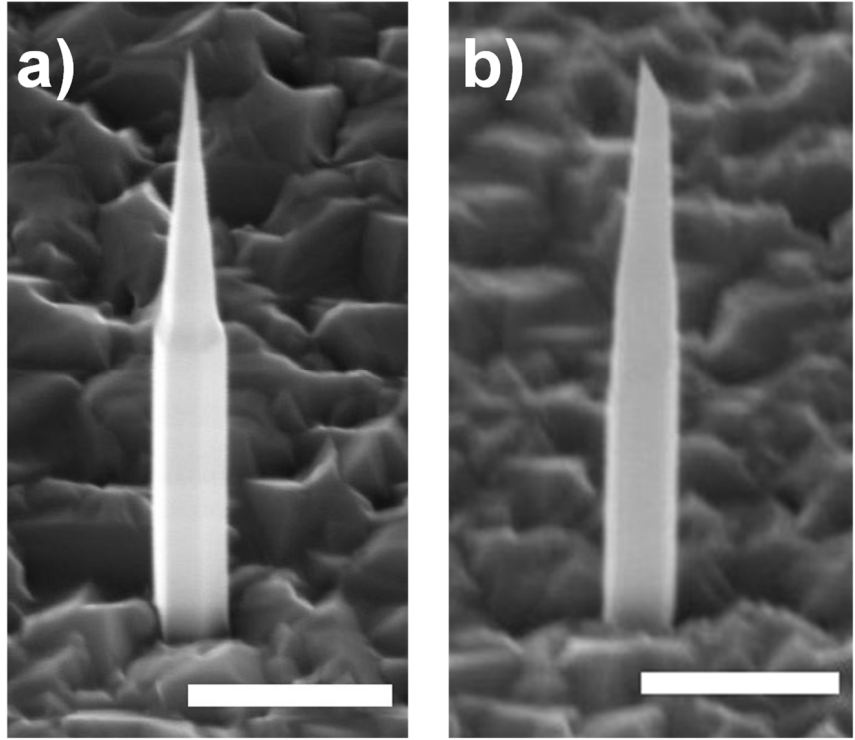

**Figure 5 –** Tilted view (45°) SEM images of (a) HS and (b) HA InAs/InP QD-NW showing the difference in the NW tapering geometry with (a) and without (b) rotating the sample during the radial growth of the InP shell. Scale bars in (a) and (b) are 1 µm.

OPTICAL SPECTROSCOPY OF HA INAS/INP QD-NWS

To study the optical polarization of the light emitted by an ensemble of InAs QDs embedded in HA InP NWs, we have performed polarization-resolved PL measurements at room temperature (T = 300 K) on A, B, C samples and compared them to a typical ensemble of HS InAs/InP QD-NWs. A 200 µm diameter laser spot size was used to investigate the PL on a QD-NW ensemble (between 500 and 800 QD-NWs are investigated). Figure 6(a-d) shows the obtained spectra from the InAs QDs embedded in (a) HS NWs and (b to d) HA NWs (A, B, and C samples), respectively. The InAs QD emission was measured at two different analyzer positions of 90° and 0° being parallel and perpendicular to the NW long cross-section axis, respectively. The integrated QD emission intensity in the 0.8 eV – 1.1 eV energy range were investigated as a function of the analyzer's angle and reported in Figure 6(e-h) for the samples with (e) HS NWs and (f to h) HA NWs (A, B and C samples), respectively.

To extract the DLP value, the obtained results are fitted by the equation:

$$I(\theta) = I_{Max}\left[\cos^2(\theta - \theta_0) + \frac{I_{Min}}{I_{Max}}\sin^2(\theta - \theta_0)\right]$$

where the $\theta$ is analyzer angle, $\theta_0$ provides the orientation angle for the maximum PL intensity and $I_{Min}$ ($I_{Max}$) is the minimum (maximum) value of the PL intensity. We also define the degree of linear polarization $DLP = (I_{Max} - I_{Min})/(I_{Max} + I_{Min})$.

As far as the symmetric NW sample is concerned a weak DLP of 4% is extracted. This confirms a negligible heavy-hole light-hole mixing for these QDs and that the PL emission from the QDs is similarly coupled to the two degenerate fundamental optical modes. This result agrees with the small height/diameter ratio of the QDs which favours a hole state with a strong heavy-hole character and with the bottom-up growth which guarantees the QD alignment with the NW central axis. This result is also in agreement with the work of van Weert et al.[26] where an average DLP of 20% was reported for vertically standing InAsP/InP QD-NWs emitting in the near infrared.

The DLP values for the InAs QDs in HA InP NWs (A, B and C samples) are 33%, 71 % and 85%, respectively. This increase of the optical anisotropy with the increase of the long/short axis ratio and the alignment of the PL polarization with the NW long axis are signatures of the impact of the anisotropic NWs on the QD emission.

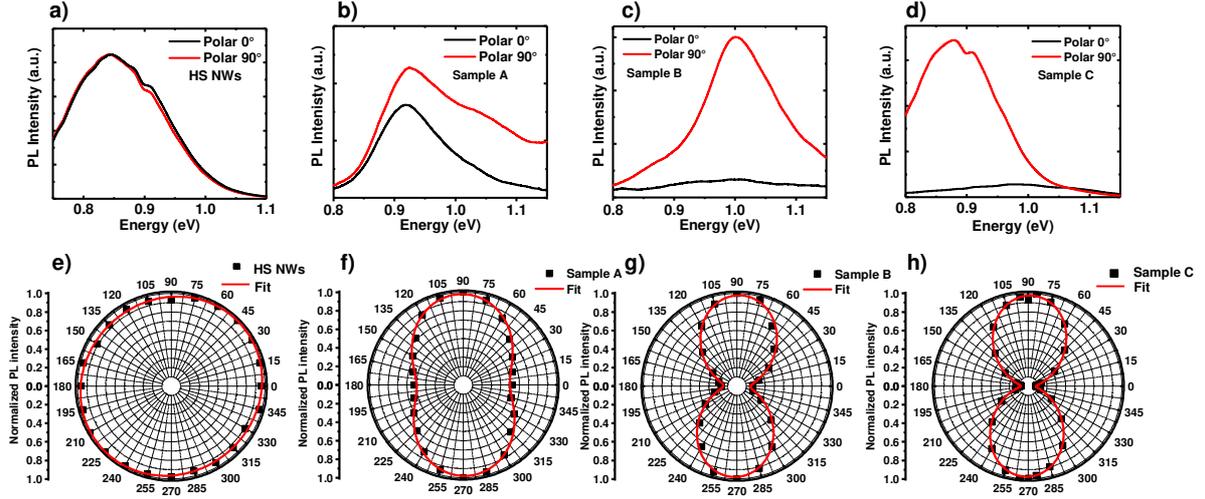

**Figure 6** – (a-d): Linear polarization-resolved PL spectra of the InAs QDs embedded in: a) HS InP NWs, b) HA InP NWs of A sample, c) HA InP NWs of B sample, d) HA InP NWs of C sample. e-h: Plots in polar coordinates of the integrated PL intensity in the 0.8-1.1 eV energy range of the spectra in (a-d) according to the analyzer angle for: e) HS QD-NWs, (f, g, h) HA QD-NWs of A, B and C samples, respectively. Solid red lines are fits to the data using the equation 1.

The DLP value as a function of the emission energy is reported in Figure 7 for the three HA QD-NW investigated samples. The highest DLP value is observed for C sample with a maximum value of 91%. This confirms the strong funneling efficiency of the QD emission with the $M_x$ mode along with the strong exclusion of the $M_y$ one in the air cladding. Moreover, a DLP ≥ 90% is observed in the 0.8 - 0.88 eV range. This broad operation bandwidth of $\Delta\lambda/\lambda=0.1$ in the telecom band presents the interest to relax the constraint on the control of the QD emission wavelength. This result agrees with previous works on symmetric and asymmetric photonic wires emitting around 950 nm where large funneling efficiencies were also reported over broad spectral ranges. A blue-shift of the maximal value of DLP is also observed in Figure 7 when the long axis of the NWs is decreased (sample C to sample A). This is explained by the importance of the $D_{NW}/\lambda$ ratio to control the funneling efficiency. In the literature, the wet chemical etching of asymmetric photonic mesa structures has been investigated to increase the DLP of InAs/InGaAlAs/InP quantum dashes (QDashes) emitting in the telecom band.[27,28]

Measurements on single QD dashes revealed a maximum DLP of 0.85.[28] Our result, DLP ≈ 0.9 at 1500 nm on a QD-NW ensemble, confirms the pertinence of asymmetric QD-NWs to reach a high polarization anisotropy.

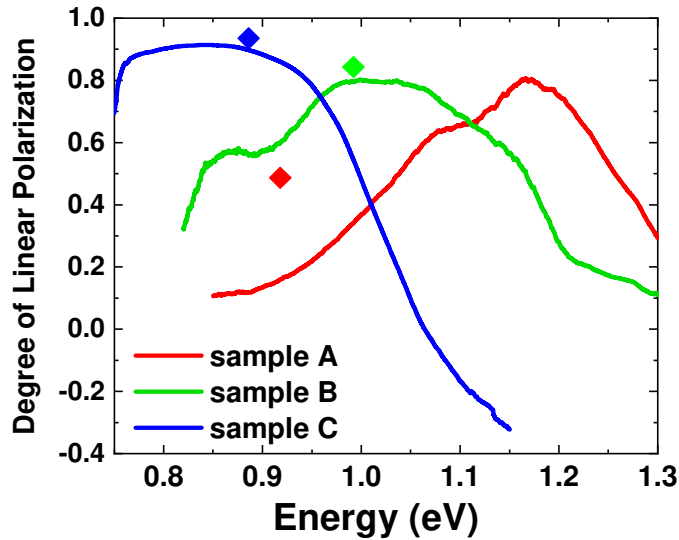

**Figure 7** - DLP values of the A, B and C samples as a function of the QD emission energy. Diamonds are the theoretical DLP values calculated using a FDTD software.

FDTD simulations were performed to corroborate the experimental results (with the use of Rsoft FullWAVE software). Theoretical values of the DLP were calculated considering the asymmetric shape of the NWs (Table 1) and the collection efficiency of the experimental setup (details in the method section). The theoretical values are reported in Figure 7. We observed that the theoretical DLP values overestimate the $I_{Max}/I_{Min}$ ratio by 50 to 110%. Despite the difference, the trend between the three samples is good: the highest values of DLP is observed for C sample and the weakest value for A sample. The difference between experimental and theoretical values could be explained by the fact that the upper part of the NWs has been simulated by a top flat interface whereas our SEM observations have revealed a non-uniform tapering morphology (see Figure 5b). This could also be explained by the fact that the PL measurements are carried out on NW ensembles and a size distribution of the short and long axis are expected or by the small difference between the ZB and WZ refractive indices of InP.[29]

CONCLUSION

In this paper, we have succeeded in controlling the polarization state of the emitted photons from InAs QDs embedded in InP NWs by changing the NW morphology from symmetric to asymmetric cross-section with long and short axes. To realize such a structure, a specific growth procedure was used during the radial growth of the InP shell, which was performed without rotating the sample while the NW [11-20] direction was aligned with the In cell flux. Our method has confirmed the impact of the HA NWs on the DLP value of the light emitted by the InAs QDs: a high DLP ≈ 91 % was reached from HA InAs/InP QD-NWs, with the QD emission being mainly polarized parallel to the NW long cross-section axis. This work stands out as a pioneering approach compared to the reported studies in literature through the growth method of HA InP-NWs with controlled InAs QD dimensions, an emission in the telecom band and monolithic growth on a Si substrate, thus opening a very promising route for polarized single photon sources. Moreover, it can be expanded for a wide range of applications such as polarized light-emitting diodes or polarized sensitive photodetectors. In the general picture, our growth protocol offers a simple and cost-effective approach covering various applications of the asymmetric NWs.

**Methods:**

**FDTD Simulations:** FDTD simulations were performed (with the use of Rsoft FullWAVE software) assuming a semi-infinite NW to remove any back reflection of the guided mode as a consequence of the bottom NW/substrate interface. The calculations were performed for the A, B and C samples. The InP NWs (InP refractive index n=3.2) are designed with an asymmetric hexagonal shape with numerical values taken from table 1. The dipole is located along the NW center axis, 0.3 µm below the flat top facet in agreement with the growth procedure. QD dipoles emitting at 1.35 µm, 1.25 µm and 1.4 µm are used for A, B and C samples, respectively. The far field emission profile is calculated as a function of the QD dipole orientation: parallel or perpendicular to the NW long axis. The numerical aperture of the experimental setup is considered to extract a theoretical DLP from the far-field intensity patterns.

**Polarization-resolved PL spectroscopy:** The QD-NWs were excited from the top using a CW 532 nm laser source with a power, P = 42 mW. The laser light was focused on the sample using a focusing lens so that the laser spot dimension on the sample is around 200 µm. The light emitted from the QDs was collected from the upper part of the sample using a Cassegrain reflective microscope objective which collects light from a 0.15 numerical aperture (NA) to a 0.4 NA. A 830 nm long-pass filter was used to cut the light emitted by the laser. Another focusing lens was added after the filter to focus the light emitted by the InAs QDs on the optical fiber coupled to a monochromator equipped with a liquid nitrogen cooled InGaAs camera. An analyzer is added in the optical path to obtain a resolved spectrum in linear polarization. The 0° angle of the analyzer corresponds to the perpendicular orientation of the analyzer with respect to the NW long cross-section axis. The emission from the InAs QDs was recorded at different polarization angles with a step of 15° in order to study the influence of the NW long/short axis dimensions on the optical polarization of the QDs. The response function of the experimental setup was registered using an unpolarized light source as a function of the analyzer angle after the sample measurements to remove any polarization dependence of the experimental setup from the experimental data.

# AUTHOR INFORMATION

**Corresponding Author:** *E-mail: nicolas.chauvin@insa-lyon.fr

**Notes:** The authors declare no competing financial interest.

## ACKNOWLEDGMENT

The authors thank the NanoLyon platform for access to equipments and J. B. Goure and C. Botella for technical assistance. The STEM studies were carried out on a microscope acquired as part of the TEMPOS project (ANR-10-EQPX-0050).

**Graphical TOC Entry**

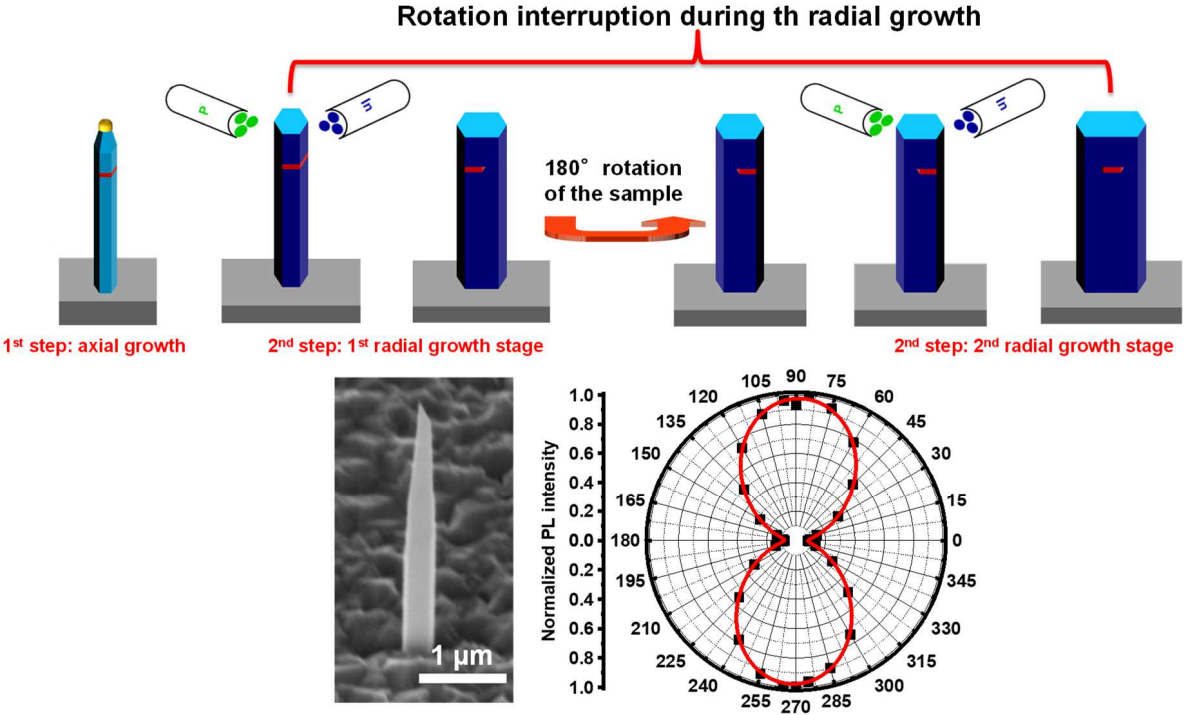